\documentclass[aps, twocolumn, superscriptaddress]{revtex4-2}

\usepackage{amsmath}
\usepackage{amssymb}
\usepackage{graphicx}
\usepackage{bm}
\usepackage{color}
\usepackage{epsfig}
\usepackage{amsfonts}
\usepackage{mathrsfs}
\usepackage{fancyhdr} 

\begin{document}

\title{On-chip  Frequency divider in superconducting quantum circuit}

\author{Hui Wang}
\email{wanghuiphy@126.com}
\affiliation{Department of Physics, Graduate School of Science, Tokyo University of Science, Shinjuku-ku, Tokyo, Japan.}
\affiliation{RIKEN Center for Quantum Computing (RQC),Wako-shi, Saitama, Japan.}

\author{Chih-Yao Shih}
\affiliation{Department of Electrophysics, National Yang Ming Chiao Tung University, Hsinchu 300093, Taiwan}
\affiliation{RIKEN Center for Quantum Computing (RQC),Wako-shi, Saitama, Japan.}

\author{Ching-Yeh Chen}
\affiliation{Department of Physics, National Tsing Hua University, Hsinchu 30013, Taiwan}
\affiliation{RIKEN Center for Quantum Computing (RQC),Wako-shi, Saitama, Japan.}

 \author{Yan-Jun Zhao}
\affiliation{Key Laboratory of Opto-electronic Technology, Ministry of Education, Beijing University of
Technology, Beijing, 100124, China}

\author{Xun-Wei Xu}
\affiliation{Key Laboratory of Low-Dimensional Quantum Structures and Quantum Control of Ministry of Education,
 Key Laboratory for Matter Microstructure and Function of Hunan Province,
  Department of Physics and Synergetic Innovation Center for Quantum Effects and Applications,
   Hunan Normal University, Changsha 410081, China}

\author{ Jaw-Shen Tsai}
\affiliation{Department of Physics, Graduate School of Science, Tokyo University of Science, Shinjuku-ku, Tokyo, Japan.}
\affiliation{RIKEN Center for Quantum Computing (RQC),Wako-shi, Saitama, Japan.}

\begin{abstract}
Based on the physical process of two-atom simultaneous excitation by single photon,
 we proposed a frequency dividing scheme in superconducting quantum circuit.
The frequency  division for a microwave photon  consists of two  quantum processes:
firstly,  two qubits share the energy of single photon in high-frequency resonator
 through the three-body interaction (two qubits and one photon); secondly, part energies of excited state qubits are
 transferred to  corresponding low frequency resonators through  two-body interactions (one qubit and one photon).
By changing the parameters of pumping pulses, controllable output pulses can  be realized through the superconducting frequency divider.
The  microwave  and pulse signals created by the superconducting frequency divider
can be used to pump or readout the superconducting qubits, which can greatly reduce the occupation amount
of high-frequency cables in dilution refrigerator  during the measurement of large scale superconducting quantum chip.
 \end{abstract}
\date{\today}

\maketitle

\section{Introduction}

As the  number of qubit increases on the superconducting quantum processors,  the  room-temperature electronic  suffers unscalable limit,
such as heating problem, linear growth of control cables,  and so on.
Many schemes have been proposed to reduce the signal noises, high-frequency cables' occupation amount and space inside the dilution refrigerator,
 such as the SFQ (single flux quantum) device\cite{McDermott,Liebermann,Opremcak,Liu,Li},
 optical control \cite{Huai,Warner} and optical readout\cite{Arnold} of qubit,
  Multiplexed superconducting qubit control at millikelvin temperatures \cite{Jiang,Acharya}.

  Digital frequency dividers,  Injection-locked dividers, Parametric dividers, MMIC (Monolithic Microwave Integrated Circuit)
  divider are serval mainstream division technology for microwave signals,  but they face challenges on phase noise degradation, limited bandwidth, power consumption, or requirements for locking over temperature and voltage variations, and so on.
The superconducting frequency divider based on  SFQ  circuit has been experimentally
realized with rather high frequency paramters and complex circuit structure\cite{Shao}.

 Inspired by recent techniques on the Integrated optical frequency
  divisions\cite{Fortier,Seeds,Yao,Hanifi,Jang,Vahala} and simultaneous excitations
  of two qubits by  single photon\cite{Tomonaga,Xin,Garziano,Yu,Bin},
   we proposed a scheme to realize the on-chip Bisection frequency divider in superconducting circuit.
 The energy of a high-frequency photon is shared by two qubits
 through the processes of  simultaneous exciting two qubits by single photon,
 after transferring the energy of  excited state qubits to  corresponding low frequency resonators,
finally a high frequency microwave photon is divided into two low frequency photons.
If we tune the parameters of  pumping pulses on the high-frequency resonator,
 the output pulse signals of different waveforms can also be realized by the superconducting frequency divider.

The proposed on-chip superconducting frequency divider takes the advantage of small sizes, low costs, and simple structure,
which can be used for the pumping and readout for superconducting qubits.
 The working mechanism of superconducting frequency divider is a pure quantum process  and should
  introduce less noises compared to the room-temperature electronic devices.
Based on the processes of single photon exciting multi-atom, two or more signal channels can be obtained by single high-frequency input signal,
thus  the  occupation amount  for high-frequency lines in the dilution refrigerator can be greatly suppressed
 during  measurement of the large scale superconducting quantum chip.

This paper is organized as follows: In Sec. II, we build the physical mode of superconducting frequency divider.
In Sec. III, the output photon number and conversion efficiency of Bisection
frequency divider are studied.  In Sec. IV, the controllable output pulse of  frequency divider ae  analyzed.
We finally summarize the results in Sec. V.

\section{Physical model}

 The  simultaneous excitation of two qubits by single photon originates
  from the third-order quantum states exchange interaction in superconducting circuit\cite{Kockum},
which can be induced by the spin-spin coupling\cite{Tomonaga,Xin},
the ultra-strong qubit-resonator coupling \cite{Garziano},
or the photon-mediated Raman process\cite{Yu}. To create a relatively
 clear anti-crossing levels for two-atom simultaneous excitations process by single photon,
 the  qubit with large anharmonicity should be  better.

\begin{figure}
\centering\includegraphics[bb=-10 120 600 460, width=8.8 cm, clip]{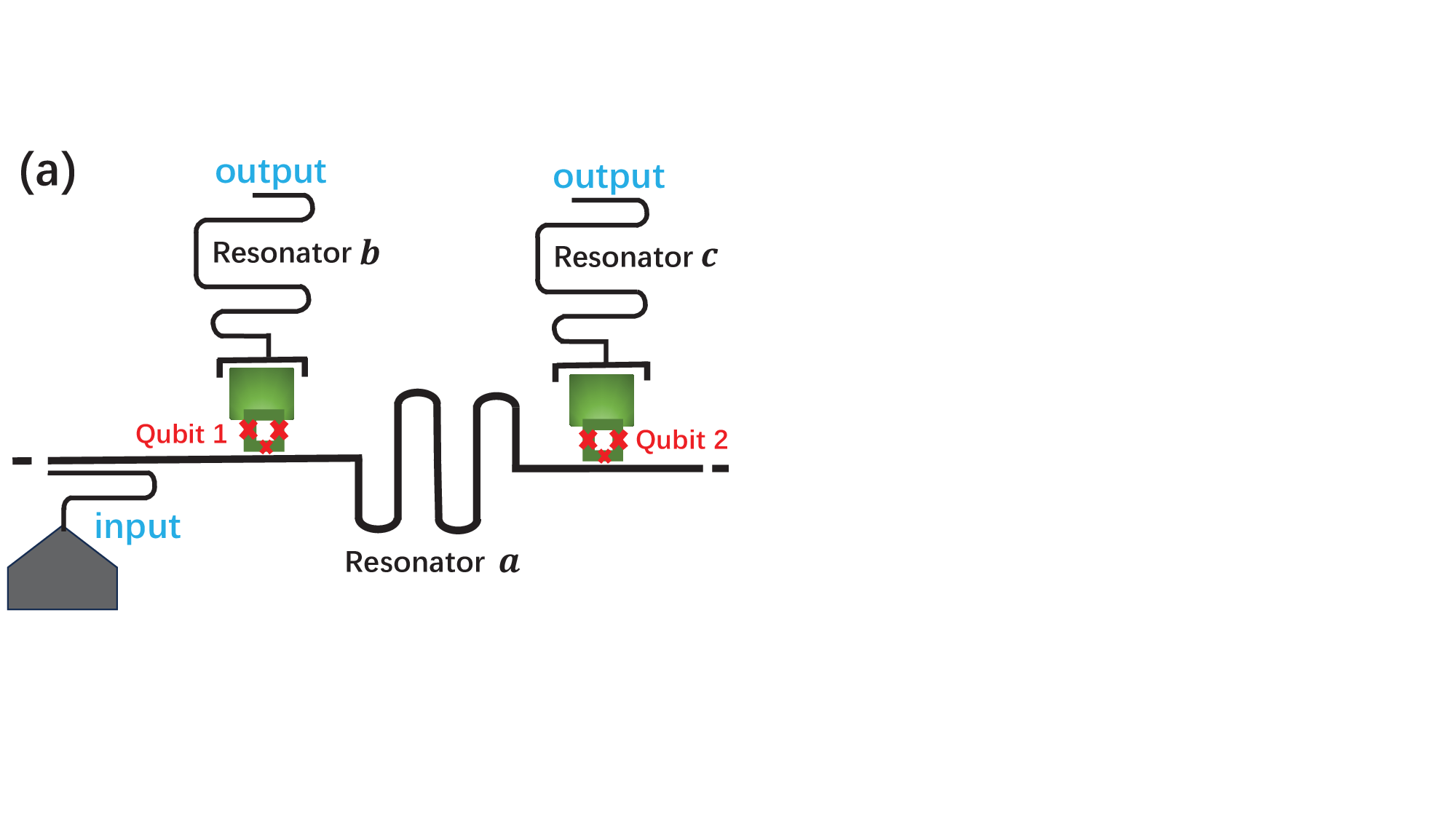}
\centering\includegraphics[bb=0 80 600 430, width=8.3 cm, clip]{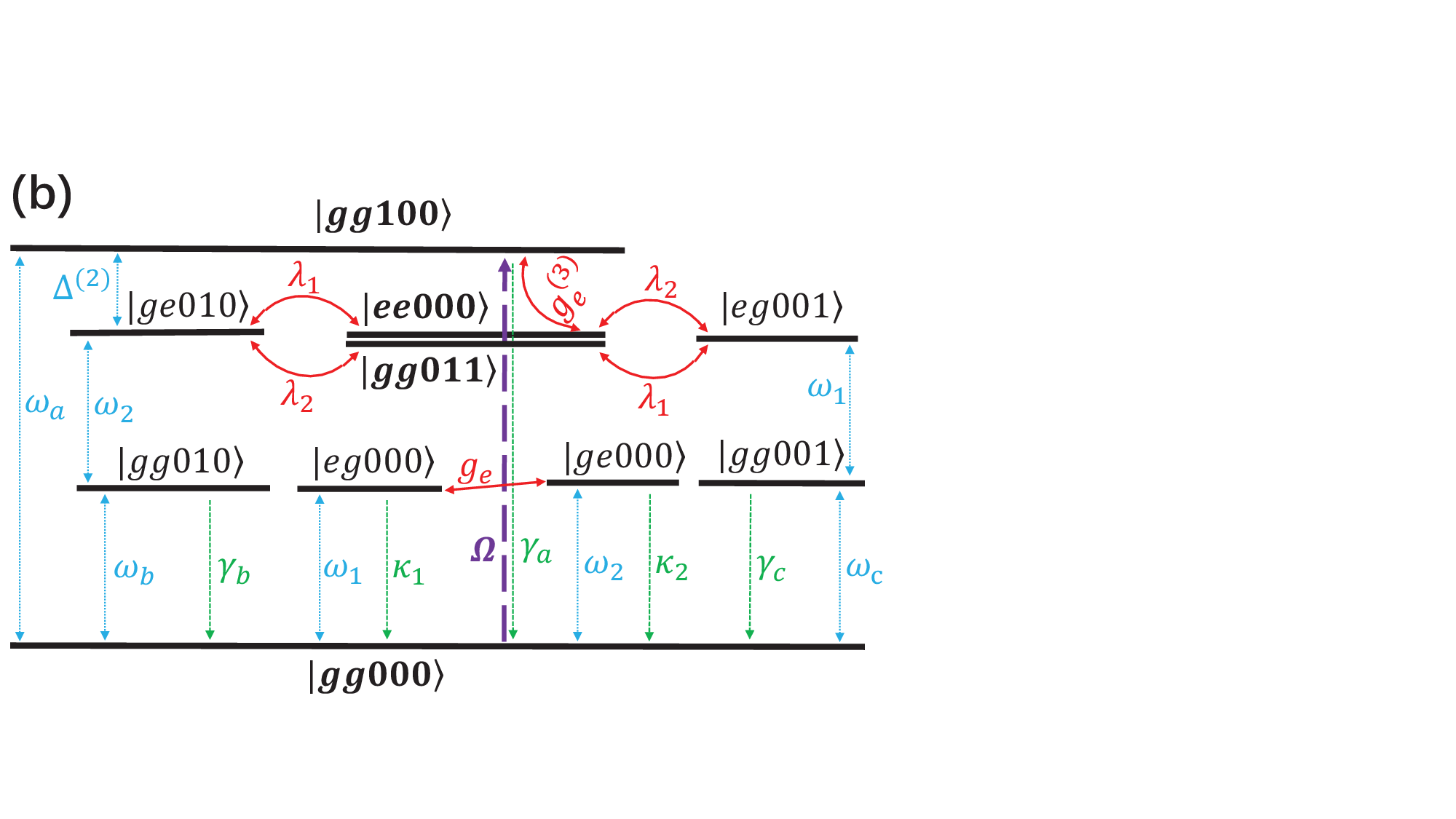}
\caption{(Color online)  (a) Sketch of  superconducting frequency divider.
The superconducting circuit consist of two three-Josephson junctions flux qubits,
two low-frequency quarter-wave resonators, and one high-frequency half-wave length resonator.
(b) Energy level structure for near resonant interactions.
The black-solid lines label the energy levels of quantum states, and energy levels for the
states $|ee000\rangle$ and $|gg011\rangle$ are near resonant.
The double-headed red-solid arrows denote exchange interactions between the energy levels,
 while the blue-dotted curves label the frequency differences of energy levels ,
and the green-dotted arrow describe the energy damping processes.  The purple dashed-arrow
describes the pumping signal with amplitudes $\Omega(t)$ and frequency $\omega_a$.
We define $\Delta^{(2)}=\omega_a-(\omega_1+\omega_2)$,
 $\omega_1=\omega_b$,   $\omega_2=\omega_c$, and $ \omega_{1} \approx \omega_{2} $.
$g_e$ labels the effective two-body coupling between two qubits, and $g^{(3)}_e$ denotes
 the effective three-body interaction among resonator-\textbf{a},
qubit-\textbf{1}, and qubit-\textbf{2}.
 }
\label{fig1}
\end{figure}

The superconducting quantum circuit in Fig.1(a) consists of two  flux qubits and three resonators.
The qubit-\textbf{1} (or qubit-\textbf{2}) capacitively  couple to low frequency resonator-\textbf{b} (or resonator-\textbf{c})\cite{Yamamoto}.
Two qubits inductively interact with a common high-frequency resonator-\textbf{a} where the longitudinal  qubit-resonator coupling makes ‌significant contributions\cite{Tomonaga,Niemczyk,Yoshihara,Liuw}.
We define $a^{\dagger}$ ($a$)  and $\sigma^{(j)}_{x,z}$ as the creation (annihilation) operator of photon in resonator-\textbf{a}
 and Pauli-{$\{X,Z\}$}  operators of qubit-\textbf{j} (with $j=1,2$), respectively.
Then the interactions between resonator-\textbf{a} and two qubits can
be described  by  the  Hamiltonian $H_i=(a^{\dagger}+a)\sum_{j=1,2}[\cos(\theta_j)\sigma^{(j)}_x+\sin(\theta_j)\sigma^{(j)}_z]$\cite{Garziano,Stassi,Niemczyk,Yoshihara}.
The value of $\theta_j$ can be
continuously tuned by changing the external magnetic flux applying on flux qubits,  the single- and two-photon transitions can coexist
 in the case of broken parity symmetry ($\theta_j\neq 0$)\cite{Niemczyk,Liuy,Liuw,Stassi,Jun}.

In the ultra-strong coupling regime between qubits and high-frequency resonator ($\omega_a\approx \omega_1+\omega_2$),
  the  corresponding effective Hamiltonian of  two qubits simultaneous excitation
   by single photon in resonator \textbf{a} can be written as \cite{Tomonaga,Xin,Garziano,Yu,Bin},
\begin{eqnarray} \label{eq:1}
H_{three}=\hbar g^{(3)}_e\left[|e,e,0,0,0\rangle\langle g,g,1,0,0|+h.c.\right],
\end{eqnarray}
where $\hbar$ is the reduced Planck constant.
 The five quantum numbers inside  state vector respectively label  qubit-\textbf{1}, qubit-\textbf{2},
 resonator-\textbf{a}, resonator-\textbf{b}, and resonator-\textbf{c}.
 $g^{(3)}_e$ labels the effective three-body interaction  among qubit-\textbf{1},  qubit-\textbf{2},
and a high-frequency photon in resonator-\textbf{a}, which is responsible for  two-qubit
   simultaneous excitation by single photon.
   $|g\rangle$ and $|e\rangle$ are respective  the ground and excited states of qubit, while $|0\rangle$ and $|1\rangle$
describe the zero- and single-photon states of resonator, respectively.

Besides the quick two-qubit transitions induced by ultra-strong coupling strengths,
the low frequency resonators-\textbf{b}  (or resonator-\textbf{c}) in  Fig.1(a)  are nearly resonant
interactions with  qubit-\textbf{1} (or qubit-\textbf{2}), thus  part energy of   excited state qubits can be transferred
  to the corresponding low-frequency resonators .
  We choose a relative weak three-body coupling strength in  this article,
 which can be realized when the ratio of qubit-resonator
 coupling strengths (high-frequency resonator and single qubit)
  reach about ten percent of qubits' transition frequencies\cite{Garziano,Bin,Niemczyk,Yoshihara}.
Considering the  large anharmonicities of flux qubits,   the  effective  Hamiltonian
 for the near-resonant transitions in the superconducting circuit of Fig.1 can be written as
\begin{eqnarray}\label{eq:2}
\frac{H_e}{\hbar} &=& \sum_{\xi=a,b,c}\omega_\xi \xi^{\dagger}\xi+\sum^{2}_{j=1} \omega_{j}\frac{\sigma^{(j)}_{z}}{2}+ \left(\lambda_1 b  \sigma^{(1)}_{+}+\lambda_2 c\sigma^{(2)}_{+}+h.c.\right)\nonumber\\
&+& g^{(3)}_e\left[a^{\dagger}\sigma^{(1)}_{-}\sigma^{(2)}_{-}+a\sigma^{(1)}_{+}\sigma^{(2)}_{+}\right]+g_{e} \sigma^{(1)}_{z}\sigma^{(2)}_{z}\nonumber\\
&+&\Omega(t)\left[ a\exp{(i\omega_a t)}+a^{\dagger}\exp{(-i\omega_a t)}\right].
\end{eqnarray}
$\Omega(t)$ and  $\omega_a$ are  respective amplitude
and frequency of pumping signal for  resonator-\textbf{a}.
 $\xi^{\dagger}$ and $\xi$  are  the respective creation and annihilation operators
of resonator-$\xi$ (with $\xi=a,b,c$), while $\sigma^{(j)}_{\pm}$ and $\sigma^{(j)}_{z}$
are the ladder  and Pauli-Z operators of qubit-$j$  (with $j=1,2$), respectively.
$\omega_{\xi}$ is the resonant frequencies of resonator-$\xi$, $\omega_j$ is the transition frequency of qubit-$j$,
 and  they satisfy $\omega_a\approx \omega_1+\omega_2$,  $ \omega_{1} \approx \omega_{2} $,
 $ \omega_{b} = \omega_{1}$, and $ \omega_{c}= \omega_{2} $.
 $\lambda_{1}$ (or $\lambda_{2}$) describes the interaction
  strength between qubit-\textbf{1} and resonator-\textbf{b} (or qubit-\textbf{2} and resonator-\textbf{c}).
 $g_e$ labels the effective coupling between first excited states of two qubits ($|ge000\rangle$ and  $|eg000\rangle$),
which consists of qubit-qubit interactions of direct and indirect types(induced by the
 resonator)\cite{Yan,WuY,Wang1,Wang2,Wang3}.
With a carefully design, the indirect and direct qubit-qubit coupling can cancel with each other ($g_e/(2\pi)\approx0$ Hz),
  thus the state mixing between two qubits are ignored.

The energy level structures for the near resonant transition of the superonducting divider can be seen in Fig.1(b).
Under the resonant pumping field,  resonator-\textbf {a} transits to the first-excited state
 ($|g,g,0,0,0\rangle\rightarrow |g,g,1,0,0\rangle $).
  After sharing the energy of a high-frequency photon in resonator-\textbf{a},
 two qubits   simultaneously transit to their first-excited states ($|g,g,1,0,0\rangle \rightarrow   |e,e,0,0,0\rangle $).
Under the continuous wave pumping,  part energy of excited state qubit can cycle back to the resonator-\textbf{a} ($|e,e,0,0,0\rangle \rightarrow  |g,g,1,0,0\rangle$).
Because of the interactions between qubits and low frequency resonator-$b$, the excited state energy of   qubit-\textbf{1} (or qubit-\textbf{2}) can transfer energy to resonator-\textbf{b} (or resonator-\textbf{c}), thus  and some energy can transfer to the low frequency resonator ($ |e,e,0,0,0\rangle \rightarrow  |g,e,0,1,0\rangle $,
$ |e,e,0,0,0\rangle \rightarrow  |e,g,0,0,1\rangle $ ). On the whole,   a high-frequency photon in resonator-\textbf{a} is  divided into two
  low frequency photons storing in  resonator-\textbf{b} and  resonator-\textbf{c}, respectively.

For the pumping signal with constant  amplitudes,
 we define the following unitary transformation to eliminate the rapidly oscillating terms for the  Hamiltonian in Eq.(2),
\begin{eqnarray} \label{eq:4}
U=\exp{\left[i\omega_a t\left(a^{\dagger}a+\frac{ b^{\dagger} b+c^{\dagger}c}{2}+\frac{1}{4}\sum^2_{j=1}\sigma^{(j)}_z\right)\right]}.\quad
\end{eqnarray}
For the single  operator, we  get
$U^{\dagger} a^{\dagger}U=a^{\dagger}\exp{(i\omega_a t)}$,
$U^{\dagger} b^{\dagger} U=b^{\dagger}\exp{\left(i\omega_a t/2 \right)}$,
$U^{\dagger} c^{\dagger} U=c^{\dagger}\exp{\left(i\omega_a t/2 \right)}$,
$U^{\dagger}\sigma^{(j)}_z U=\sigma^{(j)}_z$,
and $U^{\dagger}\sigma^{(j)}_+ U=\sigma^{(j)}_+ \exp{\left(i\omega_a t/2  \right)}$.
 And  the effective  Hamiltonian $H^{(2)}_r=U^{\dagger} H_e U-i\hbar U^{\dagger} \partial U/\partial t$,
\begin{eqnarray} \label{eq:5}
\frac{H_r}{\hbar} &=& \Delta_{ba} b^{\dagger}b+ \Delta_{ca} c^{\dagger}c+\frac{1}{2}\sum^2_{j=1}\Delta^{(j)}_{qa}\sigma^{(j)}_{z}\nonumber\\
&+ &  g^{(3)}_e\left(\sigma^{(1)}_{-}\sigma^{(2)}_{-}a^{\dagger} +a\sigma^{(1)}_{+}\sigma^{(2)}_{+}\right)\nonumber\\
&+ & \left[\lambda_1 b\sigma^{(1)}_{+}+\lambda_2 c \sigma^{(2)}_{+}+h.c.\right]+\Omega \left(a+a^{\dagger}\right).
\end{eqnarray}
We have defined $\Delta_{ba}=\omega_{b}-\omega_a/2$, $\Delta_{ca}=\omega_{c}-\omega_a/2$, and $\Delta^{(j)}_{qa}=\omega_{j}-\omega_a/2$.
  $\lambda_{1}$ (or $\lambda_{2}$)  describe the interaction of  qubit \textbf{1}-resonator \textbf{b}  (or qubit \textbf{2}-resonator \textbf{c} ).
The frequency oscillating terms are removed by the Unitary transformation,
thus the numerical calculations with the master equation should be more efficient.

If we define  $\gamma_\xi$ (with $\xi=a,b,c$) and $\kappa_{j}$ (with $j=1,2$) as  corresponding damping rates  of resonators and qubits,
 the quantum states of qubits
and photons can be calculated with the master equation\cite{Johansson,Nation},
\begin{eqnarray} \label{eq:3}
\dot{\rho}(t)=\frac{1}{i\hbar}[H_r,\rho(t)]+\sum_{o}\Gamma_{o}\left(L_{o}\rho L^{\dagger}_{o}
-\frac{1}{2} \{L_{o}L^{\dagger}_{o},\rho\}\right),\quad
\end{eqnarray}
where $\rho$ is the  density operator of   system,  with $o=a, b,c, \sigma^{(j)}_{-}$.
$L_{\xi}=\sqrt{\gamma_\xi}\xi$ and $L_{\sigma^{(j)}_{-}}=\sqrt{\kappa_{j}}\sigma^{(j)}_{-}$ for qubit-$j$ are respective
 Lindblad terms of resonator-$\xi$ and qubit-$j$. The  pure dephasing terms do not cause the qubits to lose energy or transitions,
  and they  are neglected in this article.

\begin{figure}
 \centering\includegraphics[bb=0 0 850 950, width=8 cm, clip]{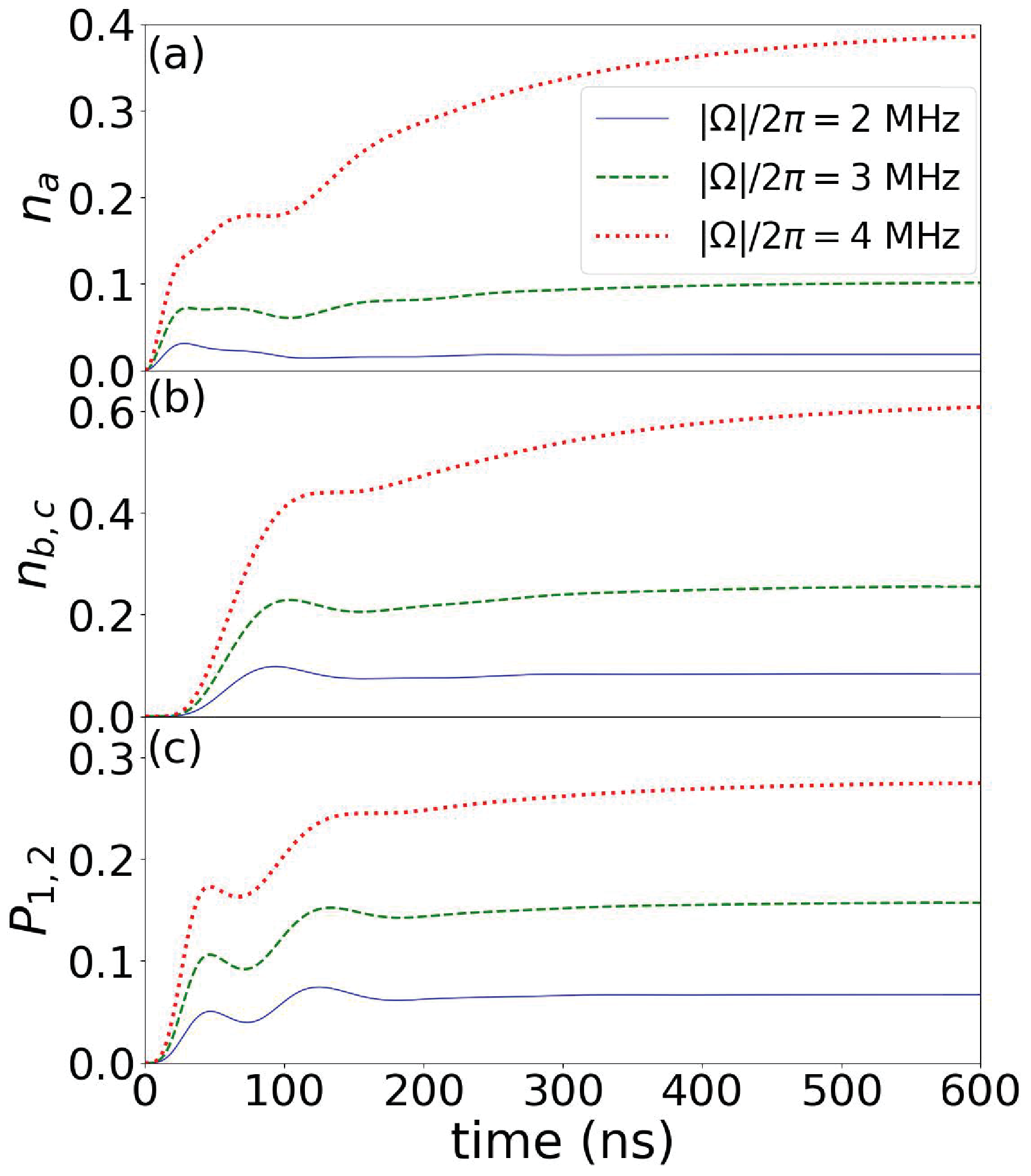}
\caption{(Color online) Output photon numbers.
(a) $n_a$, (b) $n_{b,c}$, and (c) $P_{1,2}$ as the functions of evolution time.
  The three-colored curves in  each figure of (a)-(c) for different pumping amplitudes: (1) $|\Omega|/2\pi=2$ MHz (blue-solid curve);
  (2) $|\Omega|/2\pi=3$ MHz (green-dashed curve); and (3) $|\Omega|/2\pi=4$ MHz (red-dotted curve).
   The other parameters are: $\omega_a/2\pi = 8.2$ GHz, $g^{(3)}_e/2\pi = 10$ MHz,
   $\lambda_{1,2}/2\pi = 5$ MHz, $|\Omega|/2\pi =4$ MHz, $\omega_{1,2,b,c}/2\pi=4.10$ GHz,
   $\omega_a/2\pi=8.20$ GHz, $\gamma_{a}/2\pi=6$ MHz,
 $\gamma_{b,c}/2\pi=1$ MHz, $\kappa_{1,2}/2\pi=3$ MHz,
and the photon numbers are truncated to $N^{(tr)}_{a,b,c}=6$
for the  numerical calculations.}
\label{fig2}
\end{figure}

\section{Output Photon Number}

With the experimental assessable parameters,  in this section we  analyze the output
low frequency signals of the superconducting frequency divider.
The weak continuous wave pumping field $(|\Omega|/\gamma_a <1)$ with constant amplitude  ($\Omega$) and resonant frequency ($\omega_a$)
are applied on the high-frequency resonator-\textbf{a}.
 As shown in Fig.2(a), the photon number $n_a$ on three-colored curves increase from zeroes and
finally stabilize at different values. The semiclassical motion equations of resonator-$a$
can be written as $  \dot{a}=-\gamma_{a} a-ig^{(3)}_e\sigma^{(1)}_{-}\sigma^{(2)}_{-}+i\Omega$.
After  temporarily ignoring the effects of qubits, the time dependent equation for photon number $n_a$
  can be approximately obtained as
 \begin{eqnarray} \label{eq:7}
n_a(t) \approx \frac{|\Omega|}{\gamma_a}\left[1 -\exp{(-\gamma_a t)}\right].
\end{eqnarray}
The qubits and resonators are initially in the ground states.
The finally stable values of $n_a$ should be approximately
proportional to pumping amplitude  $|\Omega|$, which roughly
coincides with the result on the three-colored curves in Fig.2(a).
Thus the strong output signals can be realized by enhancing the pumping amplitudes on resonator-$a$.

\begin{figure}
\centering\includegraphics[bb=5 5 1013 938, width=4.2751 cm, clip]{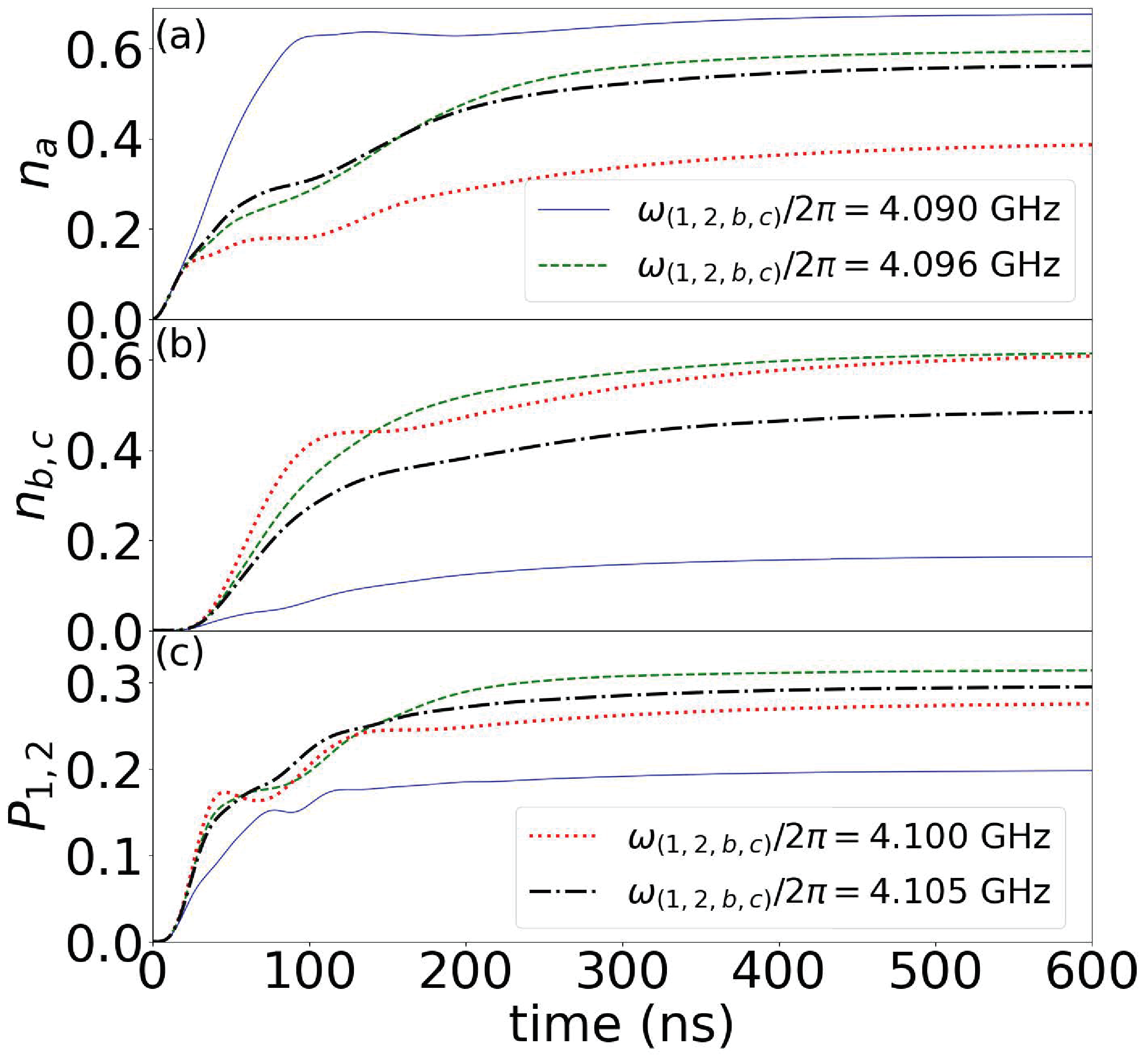}
\centering\includegraphics[bb=5 5 1012 938, width=4.2751 cm, clip]{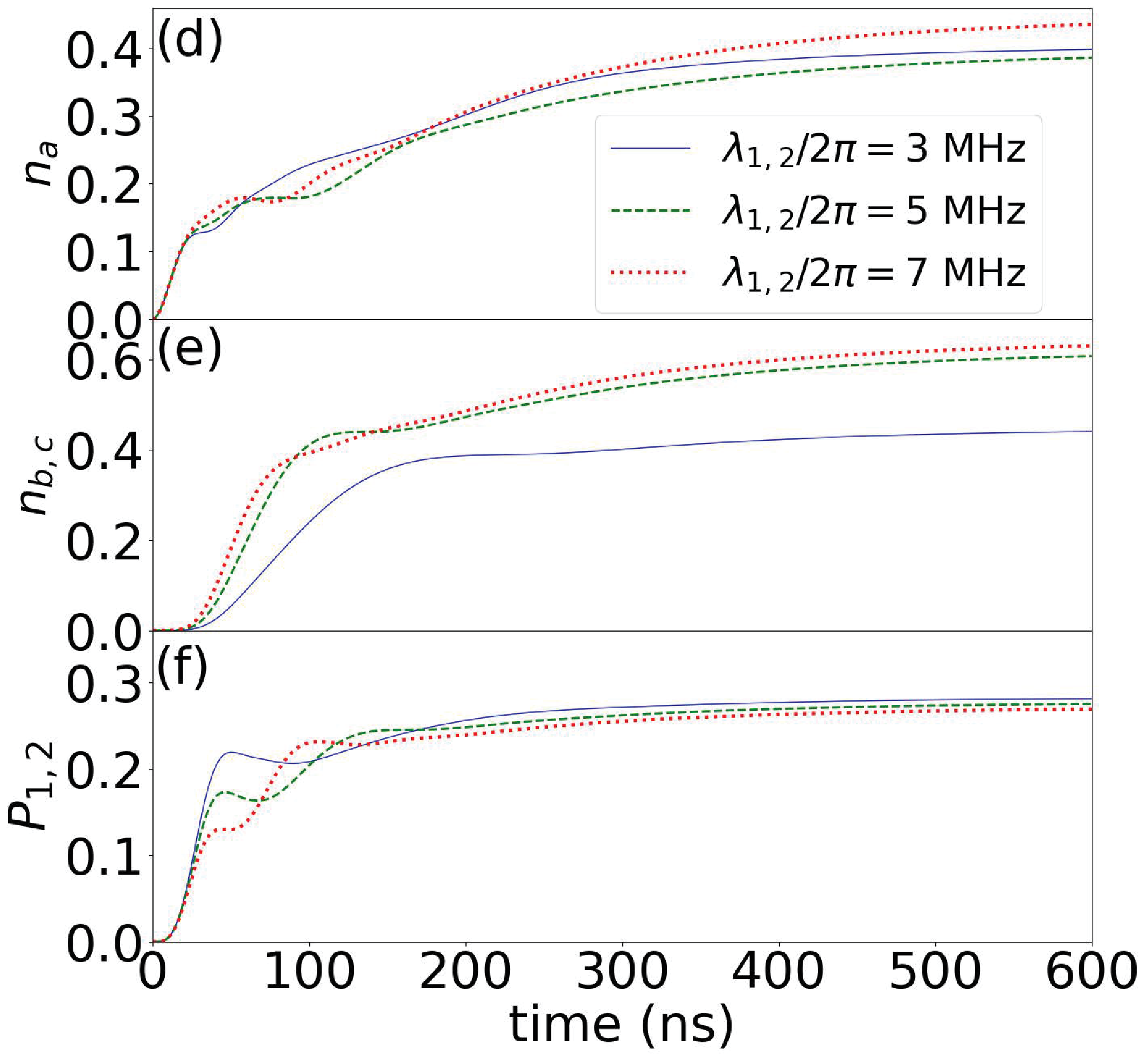}\\
\centering\includegraphics[bb=0 12 720 550, width=4.275 cm, clip]{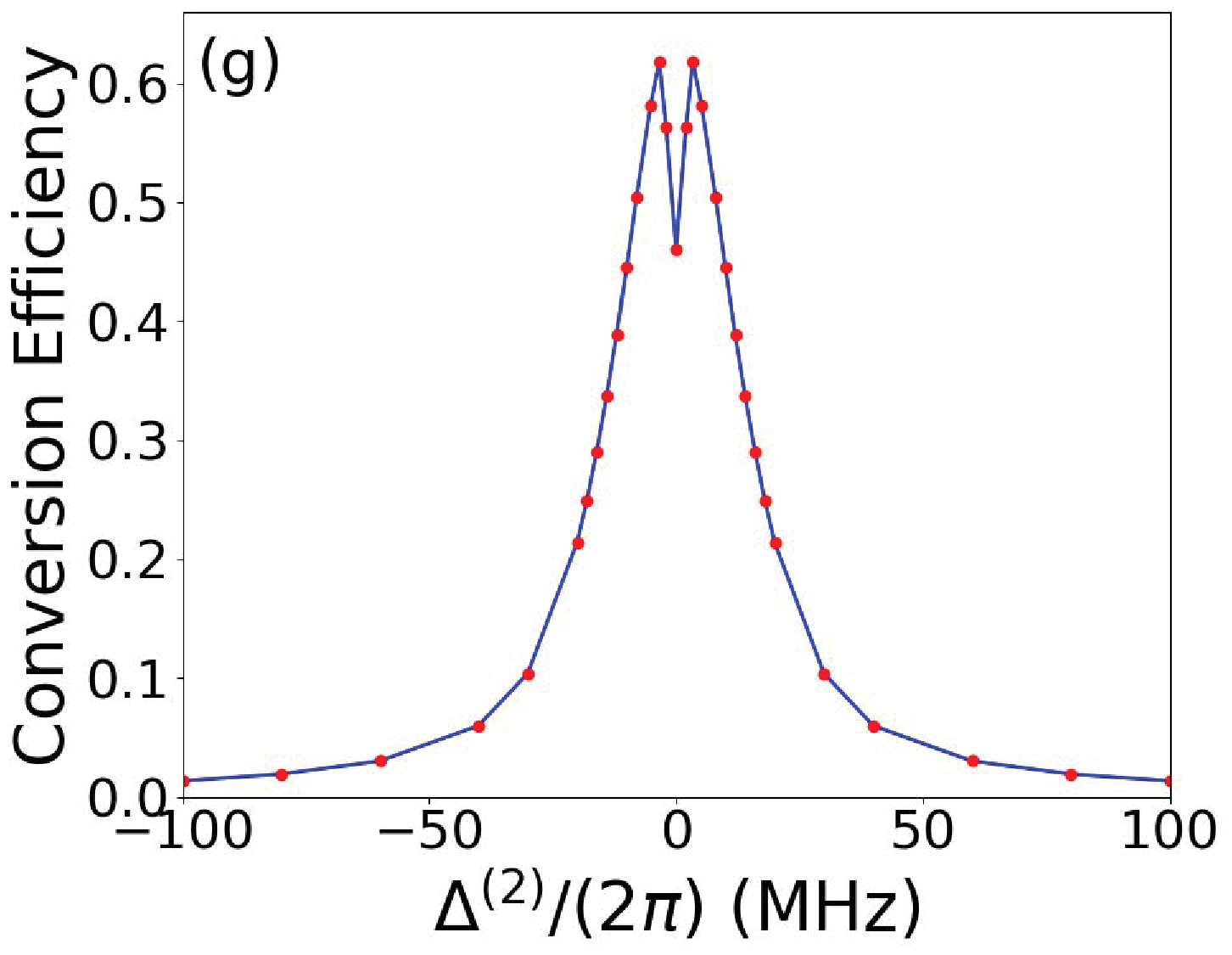}
\centering\includegraphics[bb=0 12 720 550, width=4.275 cm, clip]{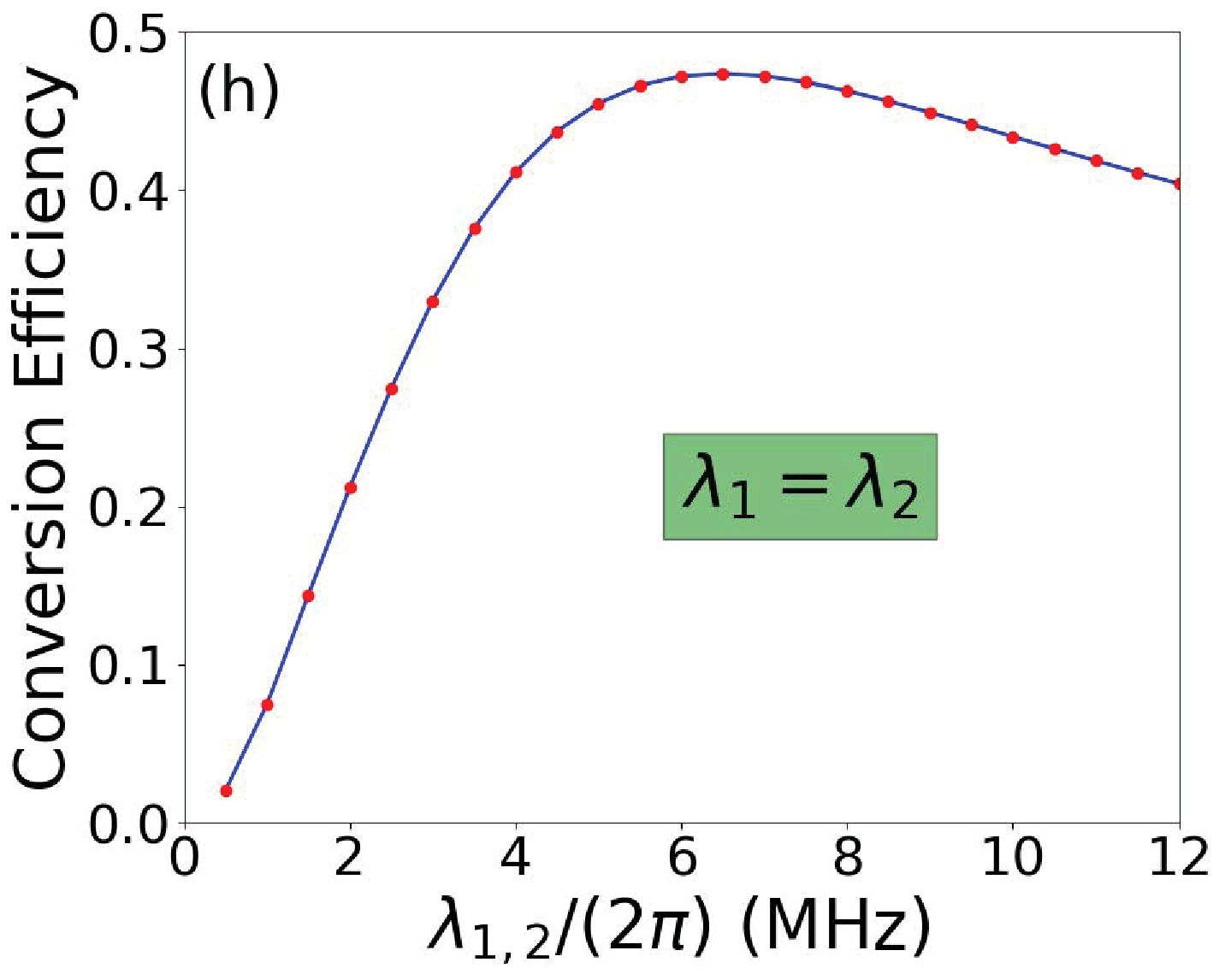}
\caption{(Color online)  Conversion efficiency.
The four-colored curves in (a) $n_a$, (b) $n_{b,c}$,  and (c) $P_{1,2}$
correspond to different frequency detuning ($\Delta^{(2)}=\omega_a-(\omega_1+\omega_2)$),
with $\omega_{1,2,b,c}/2\pi=$: (1) 4.090 GHz (blue-solid curve);(2) 4.096 GHz (green-dashed curve);
  (3)4.100 GHz (red-dotted curves); and (4) 4.105 GHz (black-dash-dotted curve).
  The three-colored curves in  (d) $n_a$, (e) $n_{b,c}$,  and (f) $P_{1,2}$
  for different qubit-resonator coupling strengths: (1) $\lambda_{1,2}/2\pi=3$ MHz (blue-solid curve);
  (2)$\lambda_{1,2}/2\pi=5$ MHz (green-dashed curve); and (3) $\lambda_{1,2}/2\pi=7$ MHz (red-dotted curve).
    Conversion efficiency of   frequency divider
     as the functions of (g) frequency detuning $\Delta^{(2)}$
      and (h) coupling strength $\lambda_{1,2}$.
  The two qubits are initially in the ground states,
   and the other parameters are: $\omega_{a}/2\pi=8.20$ GHz,
  $\omega_{1,2,b,c}/2\pi=4.10$ GHz ((d)-(f),(h)), $g^{(3)}_e/2\pi = 10$ MHz,
  $\lambda_{1,2}/2\pi = 5$ MHz ((a)-(c),(g)), $\gamma_a/2\pi=6$ MHz, $\gamma_{b,c}/2\pi=1$ MHz,
  $\kappa_{1,2}/2\pi=3$ MHz, $|\Omega|/2\pi=4$ MHz , and the photon numbers are truncated
   to $N^{(tr)}_{a,b,c}=6$.}
\label{fig3}
\end{figure}

 In the  parameter regimes  of $\omega_a\approx \omega_1+\omega_2$,
 the  state exchange  induced by the  three-body interaction in Eq.(1)  can lead to the system cycling
 back and forth between single photon excited state ($|g,g,1,0,0 \rangle$) and two qubits excited state ($|e,e,0,0,0\rangle$), which results in the oscillation
   on the  curves of $n_a(t)$ and $P_{1,2}(t)$ in the short time scale ($t<1/g^{(3)}_e$).
Similar tendencies also appear on the evolution curves of low-frequency photon number $n_{b,c}$
in Fig.2(b) and  qubits' occupation probabilities ($P_{1,2}$) in Fig.2(c).
In the case of same parameters for qubits or low frequency resonators,
the occupation probabilities of qubits ($P_{1,2}$)  or low frequency  photon numbers ($n_{b,c}$) should be the same,
 so they are described by the same  curves in this article.
Thus the  on-chip superconducting frequency divider can split an input high-frequency microwave signals of 8.2 GHz   into two low frequency signals of frequency 4.1 GHz.
 If  two qubits are designed with non-identical transition frequencies, the output signals with different  frequencies can be realized.

The output low frequency signals can be measured by detecting the photon number inside the low frequency resonators.
The variations for the  resonator's intrinsic quality factor\cite{Tominaga} or
dephasing rate of  qubit (Off-resonant  interaction with resonator) \cite{Atalaya}
  can be used to detect the photon number in the low-frequency resonators.
The  vacuum Rabi couplings can be observed during the  energy exchange processes between qubit and  resonator
\cite{Mariantoni}, which can also be used to discern the photon number inside the low-frequency resonators.

\subsection{Conversion Efficiency}

In this section, we study the effects of frequency detuning ($\Delta^{(2)}=\omega_a-(\omega_1+\omega_2)$)
 and  qubit-resonator coupling strengths ($\lambda_{1,2}$)
 on the output low-frequency photon number and conversion efficiency of superconducting  frequency divider.
The frequency of superconducting devices  can't be  identical in experiment, and the dependent relation
of output low-frequency photon numbers ($n_{b,c}$) on the frequency detuning ($\Delta^{(2)}$)  are shown in Fig3(a).

In the case of $\Delta^{(2)}/(2\pi)=20$ MHz,
the large frequency detuning reduces the efficiency of two qubit simultaneous excitation.
Thus $n_a$  get the highest values in the  stable regimes of blue-solid curve in Fig.3(a) , while
 the output photon numbers $n_{b,c}$ and qubits' occupation probabilities $P_{1,2}$
 get lowest values on the  blue-solid curves of Figs.3(b) and 3(c), respectively.
For $\Delta^{(2)}/(2\pi)=8$ MHz,    $n_{b,c}$  in Fig.3(b) and  $P_{1,2}$ in Fig.3(c)
get the largest values on  the green-dashed curves (stable regimes),
and they are even larger than the result  on the red-dotted curves for the zero-detuning case ($\Delta^{(2)}/(2\pi)=0$ Hz).
The near-resonant coupling between qubits and corresponding low frequency resonators create    dressed states,
which  shifts the qubits' energy levels and add additional frequency detuning for two qubits simultaneous excitations.
  The variations of output photon numbers as the functions of  frequency detuning
$\Delta^{(2)}$ reveals the bandwidth of the superconducting frequency divider,
 which can be enhanced by changing the damping rates of resonators.

In this article, we only focus on the  conversion  efficiency of the frequency divider in stable regimes.
According to the input-output relation \cite{Gardiner,Zou,Hill}, the conversion  efficiency of an
 input high-frequency  photon into two output low-frequency photons in  superconducting
frequency divider can be defined as
\begin{eqnarray}
T=\left|\frac{\gamma_b n^{(s)}_b \omega_b+ \gamma_c n^{(s)}_c\omega_c}{\omega_a|\Omega|^2/(2\gamma_a)}\right|,
\end{eqnarray}
where $n^{(s)}_b$ and  $n^{(s)}_c$ are stable photon numbers in resonators $b$ and $c$, respectively.
The input power pumping field to resonator-$a$ is defined as  $P_{in}=\hbar\omega_a|\Omega|^2/(2\gamma_a )$.
The internal losses of resonators are assumed to be much smaller than the external losses.

\begin{figure}
\centering\includegraphics[bb=0 74 600 410, width=8 cm, clip]{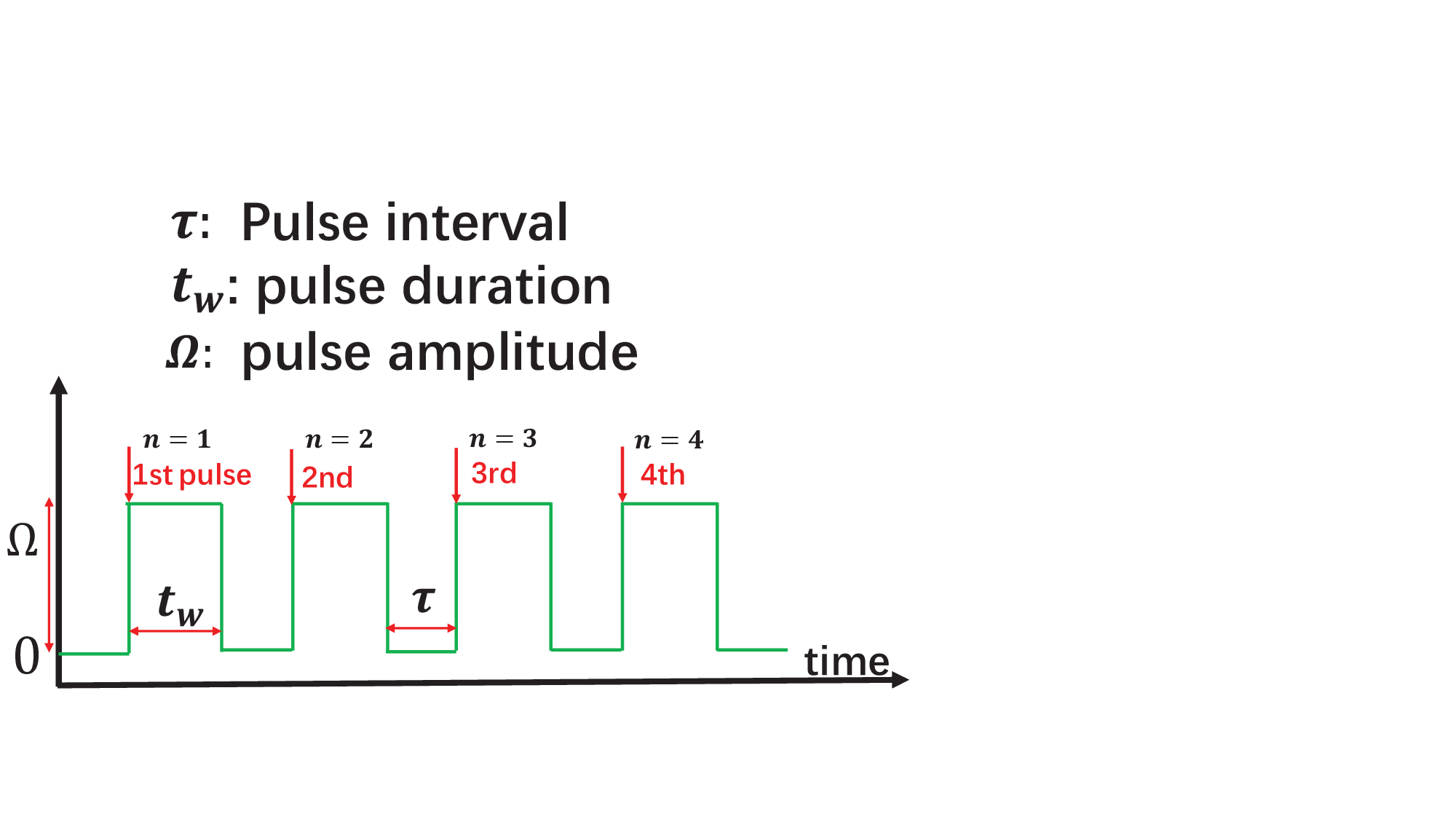}
\caption{(Color online) Square wave pulse.
$t_w$ denotes the  pulse duration (pulse width),
 $\tau$ labels the pulse interval, while $\Omega$
is the  pulse amplitude.}
\label{fig4}
\end{figure}

 In the  stable regimes of output photon number $n_{b,c}$,   the conversion efficiency
  as the function of frequency detuning $\Delta^{(2)}$ is calculated as shown in Fig.3(g).
The highest conversion efficiency  ($\max{(T)}$) locates at the two peaks around the zero detuning point ($\Delta^{(2)}/(2\pi)=0$ Hz),
which results from the  dressed states  formed by   qubits and corresponding low-frequency resonators.
The dressed states energy levels  change the conditions of two qubit simultaneous excitation, thus
 the relatively larger output photon number appears on the green-dashed curve ($\Delta^{(2)}/(2\pi)=8$ MHz) in Fig.3(b).
  For the large frequency detuning regimes, such as $\Delta^{(2)}/(2\pi)>50$ MHz, the conversion efficiency
  of frequency divider becomes negligibly small as  shown in  Fig.3(g).

 The coupling strength $\lambda_{1,2}$ directly dominates the efficiency of state
  exchanging between  qubits and corresponding low frequency resonators, and it also indirectly
 affect the two qubits simultaneous excitation processes through  splitting of  dressed states energy levels.
  The three-colored curves in each images of  Figs.3(d)-3(f) correspond to different coupling strength $\lambda_{1,2}$,
  and  the effects of $\lambda_{1,2}$ on the output photon number $n_{b,c}$ are  complex as shown in Fig.3(b).
   As the variations of $\lambda_{1,2}$,  the state exchanges for single qubit transition
 and  two qubit simultaneous transition are both affected, the competitions between these two processes  decide the
  output low frequency  photon numbers.
   As indicated by  Fig.3(h), the highest conversion efficiency  locate close to $\lambda_{1,2}/(2\pi)=6$ MHz,
 this coincides with the  largest photon number $n_{b,c}$ (in stable regimes) on the green-dashed  curve in Fig.3(b).

The  conversion efficiency shown in Figs.3(g) and 3(h) did not consist of the energy leakages
to the neighbour energy levels and  high-excited states of qubits,
 thus the real conversion efficiency of superconducting frequency divider should be lower.
According to current architecture of  superconducting quantum chips,
 the  microwave signals for the pumping or readout of superconducting qubit distribute in a narrow frequency ranges.
Thus the proposed superconducting frequency divider still has wide range of applications
in the measurement of large size superconducting quantum chip despite with  a relative narrow bandwidth.

\begin{figure}
 \centering\includegraphics[bb=0 10 900 660, width=7 cm, clip]{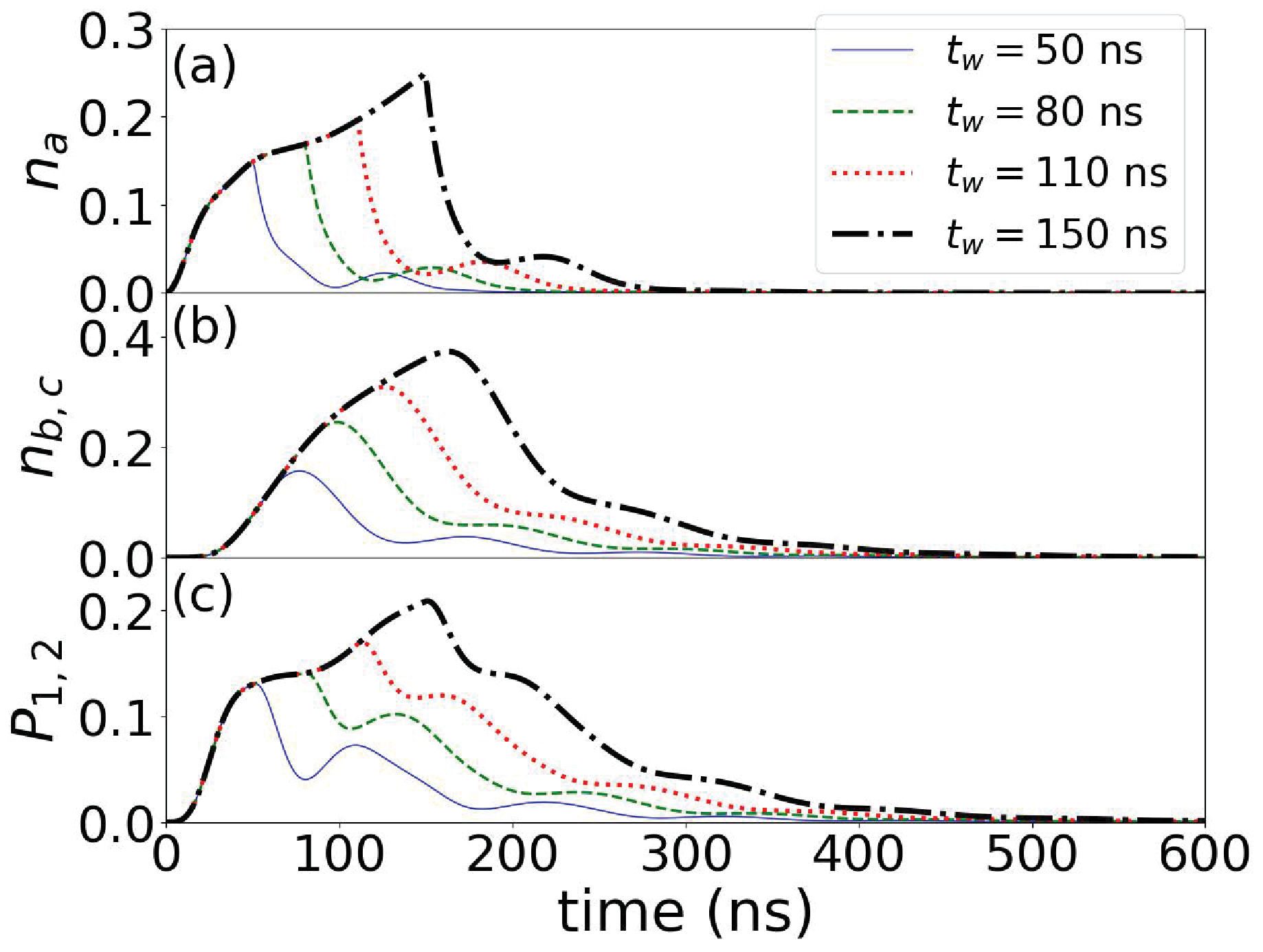}
\caption{(Color online) Pulses duration .
 The (a) $n_a$, (b) $n_{b,c}$, and (c) $P_{1,2}$ as the function of first pulse duration ($n=1$ in Fig.4).
  The four-colored curves in each figure: (1) $t_w=50$ ns (blue-solid curve);
  (2) $t_w=80$ ns (green-dashed curve); (3)  $t_w=110$ ns (red-dotted curve);
  (4)  $t_w=150$ ns (black-dash-dotted curve).
 The other parameters are: $\tau=0$ s, $\omega_a/2\pi = 8.2$ GHz, $g^{(3)}_e/2\pi = 10$ MHz,
   $\lambda_{1,2}/2\pi = 5$ MHz,  $\omega_{1,2,b,c}/2\pi=4.096$ GHz,
   $\gamma_{a}/2\pi=6$ MHz,
 $\gamma_{b,c}/2\pi=1$ MHz, $\kappa_{1,2}/2\pi=3$ MHz,  and $|\Omega|/2\pi=7$ MHz.
The photon numbers are truncated to  $N^{(tr)}_{a,b,c}=6$.
}
\label{fig5}
\end{figure}

\section{Controllable Output Pulse}

SFQ (Single-flux-quantum)-based digital signal processing is considered as a promising platform for the control of large-scale quantum processors\cite{Leonard,McDermott,Ballard,Howe,Gao}. In this section, we will try to control the microwave pulse
signals with the superconducting frequency divider.

Under the continued  pumping field,  part energy of excited state of  qubits could transfer back to
high-frequency resonator ($|e,e,0,0,0\rangle\rightarrow|g,g,1,0,0 \rangle$).
If the pumping signals maintain  certain durations, the probabilities of qubits' excited state energy
 circling back to high frequency resonator can be  suppressed, which might enhance the  efficiency of frequency divider.

For the square wave signals, the  time-dependent  amplitudes of pumping pulse can be written as
\begin{equation}
\Omega(t)=\left\{\begin{array}{l}
 \hbox{$|\Omega| \exp{(-i\omega_a t)}$,} ~ (n-1)t_p+\tau\leq t \leq  n t_p,\\
~0,   ~~~~~~~~~~~~~~~~~~~ otherwise,
\end{array}\right.
\end{equation}
where $t_p=t_w+\tau$. $t_w$ is the duration of square pulse, $\tau$ is the pulse interval,
 and $n=1,2,3,\cdot\cdot\cdot$. Substituting the pulse amplitude $\Omega(t)$ into Eqs.(4) and (5),
   the  quantum states of qubits and resonators can be numerically calculated with the master equation.

 \begin{figure}
 \centering\includegraphics[bb=0 10 900 680, width=8 cm, clip]{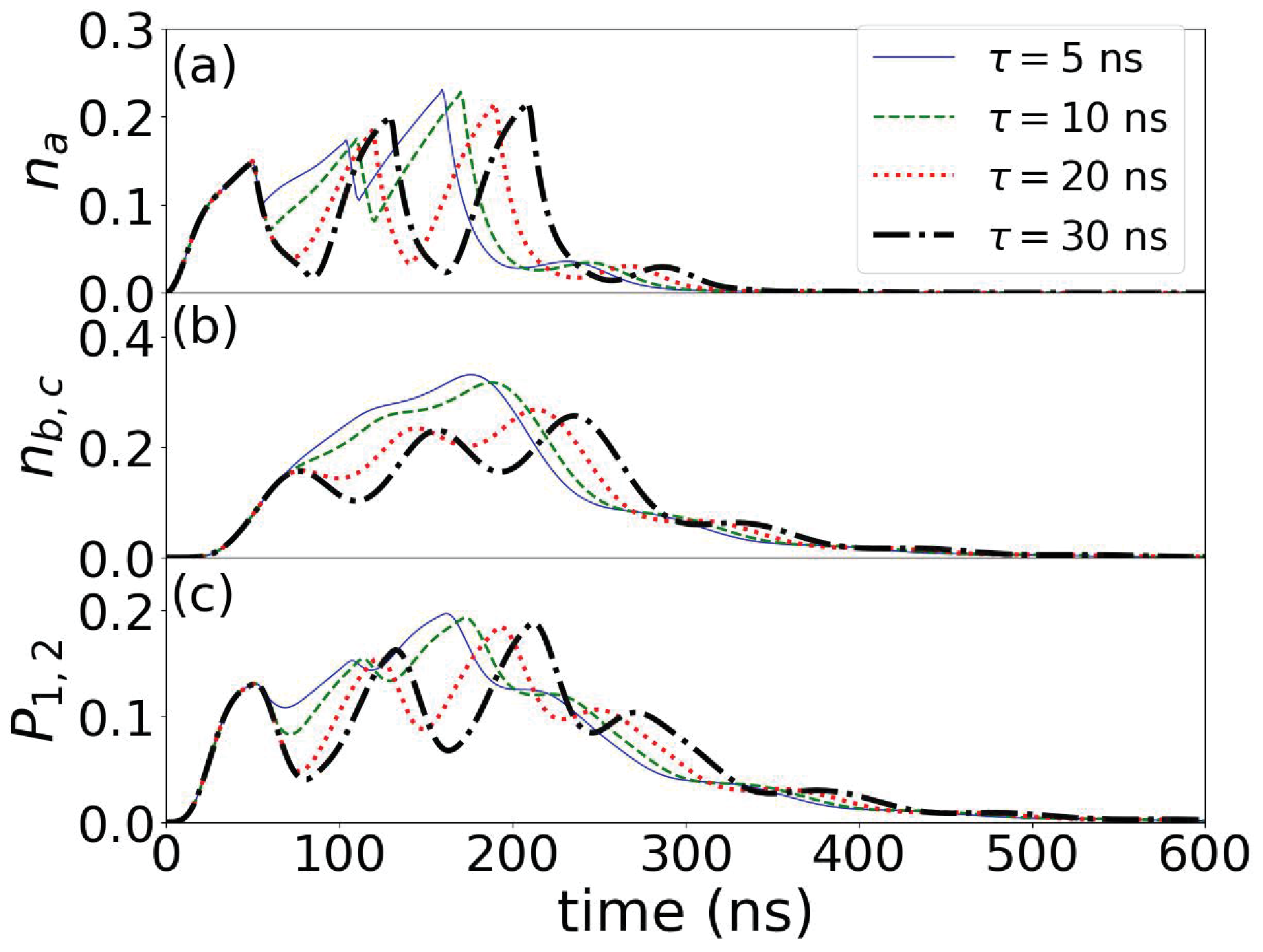}
\caption{(Color online) Pulse interval  ($n=3$ in fig.4).
 The (a)$n_a$, (b) $n_{b,c}$, and (c) $P_{1,2}$  as the function of  pulse interval (n=3).
  The four-colored curves in each figure: (1) $\tau=5$ ns (blue-solid curve);
  (2) $t_p=10$ ns (green-dashed curve); and (3)  $t_p=20$ ns (red-dotted curve),
  and (4)  $t_p=30$ ns (black-dash-dotted curve).
 The other parameters are: $\omega_a/2\pi = 8.2$ GHz, $g^{(3)}_e/2\pi = 10$ MHz,
   $\lambda_{1,2}/2\pi = 5$ MHz,  $\omega_{1,2,b,c}/2\pi=4.096$ GHz,
   $\gamma_{a}/2\pi=6$ MHz,
 $\gamma_{b,c}/2\pi=1$ MHz, $\kappa_{1,2}/2\pi=3$ MHz, $t_w=50$ns,  and $|\Omega|/2\pi=7$ MHz.
The photon numbers are truncated to $N^{(tr)}_{a,b,c}=6$.
}
\label{fig7}
\end{figure}

 By applying the first square wave pulse ($n=1$ in Fig.4) to resonator-\textbf{a},
the effects of pulse duration (width) on the frequency divider are shown in Fig.5, and the
 four-colored curves in each figure corresponds to different pulse durations.
Because of the existence of ordered for several excitation processes,
 the  rising and falling trends on the curves of  $n_a$, $n_{b,c}$, and $P_{1,2}$ are
  out of sync and sequentially increase after   applying  pulses on resonator-\textbf{a} as shown in Figs.5(a)-5(c).
  As the increase of pulse duration ($t_w$), the maximal output photon numbers $n_{b,c}$ enhance,
   but the variations of the curves  finally  level off  because of the balance between the qubits' excitation and damping losses.

By setting $n=3$ in Fig.4, a pulse  sequence consisting of three  periodic square shapes is applied on the high-frequency resonator.
The pulse duration  is fixed at $t_w=50$ns, and the pulse intervals ($\tau$) are different on the three-colored curves in each image of Fig.6.
As the variation of pulse interval, the output waveform of signal changes. In the case of $\tau=30$ns,
three separated peaks appear  on the black-dash-dotted curves of
the low-frequency photon numbers $n_{b,c}$ in Fig.6(b) and qubits' occupation probabilities  $P_{1,2}$ in Fig.6(c).
For $\tau=20$ns, the three peaks on the red-dotted curves of Figs.6(a)-6(c) becomes  gentle relative to the black-dash-dotted curves.
If the pulse interval decreases to $\tau=5$ns (or $\tau=10$ns), the peaks on the  blue-solid  curve (or green-dashed curves) almost merges as one as shown in  Fig.6(b).
 Thus it is possible to create  pulse signals through the superconducting frequency divider, and the waveform of output pulse signals are tunable.

\section{Conclusions}\label{conclusion}

Based on  the two-qubit simultaneous excitation processes in the superconducting   circuit,
 we  equally divide a high frequency photon into two low frequency photons.
If three  qubits are simultaneously excited by single photon, thus a Trisection frequency divider can also be realized in the superconducting quantum circuit.
Our proposed on-chip superconducting frequency divider scheme supply a platform to create the low frequency microwave  and pulse signals in low temperature,
the low noises signal and small occupation for the  dilution refrigerator should be useful for the measurement of  large scale superconducting quantum chip.

\section{ACKNOWLEDGMENTS}

We thank Zhiguang Yan and  Rui Wang for their valuable suggestions.
Y.J.Z. is supported by National Natural Science Foundation of China (Grant No. 62474012),
Beijing Natural Science Foundation (Grant No. 4222064), financial support from China Scholarship Council,
 Beijing Outstanding Young Scientist Program (JWZQ20240102009). X.-W.X. is supported
  by the National Natural Science Foundation of China (Grant Nos.12064010 and 12247105),
   the Science and Technology Innovation Program of Hunan Province (Grant No. 2022RC1203),
    the Natural Science Foundation of Hunan Province of China (Grant No.2021JJ20036),
     and the Hunan provincial major sci-tech program(Grant No. 2023ZJ1010).
J.S. Tsai is supported by the Japan Science and Technology
Agency (Moonshot R$\&$D, JPMJMS2067; CREST, JPMJCR1676) and the
New Energy and Industrial Technology Development Organization
(NEDO, JPNP16007).

\end{document}